\documentclass[aps,prb,twocolumn,groupedaddress,showpacs]{revtex4}

\usepackage{latexsym}

\usepackage{graphics}

\begin{document}

\preprint{p01feb02}

\title{Emergent Relativity}

\author{R. B. Laughlin}

\affiliation{Department of Physics, Stanford University,
             Stanford, California 94305}

\homepage[R. B. Laughlin: ]{http://large.stanford.edu}

\date{\today}

\begin{abstract}

A possible resolution of the incompatibility of quantum mechanics and
general relativity is that the relativity principle is emergent. I show
that the central paradox of black holes also occurs at a liquid-vapor
critical surface of a bose condensate but is resolved there by the
phenomenon of quantum criticality.  I propose that real black holes are
actually phase boundaries of the vacuum analogous to this, and that the
Einstein field equations simply fail at the event horizon the way quantum
hydrodynamics fails at a critical surface.  This can occur without
violating classical general relativity anywhere experimentally accessible
to external observers.  Since the low-energy effects that occur at
critical points are universal, it is possible to make concrete
experimental predictions about such surfaces without knowing much, if
anything about the true underlying equations. Many of these predictions
are different from accepted views about black holes - in particular the
absence of Hawking radiation and the possible transparency of cosmological
black hole surfaces. [To appear in the C. N Yang Festschrift (World 
Sci., Singapore, 2003).]

\end{abstract}


\maketitle

\section{Introduction}

It is a great honor for me to be speaking at this symposium for Prof. C.
N. Yang.  Like many other physicists, I have always envied Prof. Yang's
many excellent contributions to science over the years, and have even
shared the common experience of aspiring to match them and managing to
fall short. Dealing with this is not a happy thing.  I confess having
become depressed over it for a time and cheering up only after realizing
that everyone else had the same problem.  I now no longer worry about it.  
One should no more agonize over this inadequacy than over being too short
or bald. I recommend this course of action for the rest of you sufferers,
incidentally, in case you have not figured it out already for yourselves.  
I also recommend that we keep trying, for Prof. Yang continues to be the
man to beat.

My views on the great unsolved questions at the core of modern
physics---quantum measurement, the emergence of the correspondence limit
through decoherence, spontaneous ordering, hierarchies of laws---are
strongly influenced by my life in condensed matter physics, where
theoretical ideas are forced to immediate and brutal confrontation with
experiment by virtue of the latter's low cost. Anyone subjected to this
long enough eventually develops the habit of thinking experimentally, of
choosing experimental issues primarily on the basis of what one could
measure in a given situation, and evaluating theories mainly on the basis
of the experiments they correctly predict.  This is considered overly
conservative in many circles, but I disagree. I believe that physics is an
experimental science, and that theory acquires authority by confronting
and conforming to experiment, not the other way around.  Dealing with a
rich experimental record day after day has the additional benefit of
giving one a healthy respect for the natural world's ability to surprise
and a healthy {\it dis}respect for the belief that all things can be
calculated from first principles.

\section{The Relativity Principle}

I wish today to discuss the black hole horizon paradox and the
incompatibility of relativity and quantum mechanics.  This is obviously a
great problem in the physics pantheon and something of great interest
to all of us, particularly in light of recent advances in string theory.  
However, what I have to say is not so friendly to microscopic approaches
of this kind. I have become increasingly concerned that the essence
of the problem may not be microscopic at all but collective, and that
studying microscopic models of the vacuum may be the wrong thing to do
{\it even if the models are right}.  I think black hole formation may be a
quantum phase transition \cite{david,helium}.

\begin{figure}
\includegraphics{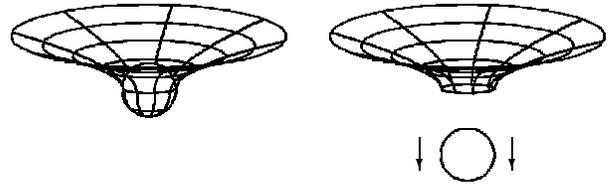}
\caption{Einstein gravity is similar to a heavy ball placed on a rubber
         membrane, except that the membrane ruptures if the ball is too
         heavy. Rupturing can be prevented by declaring the
         laws of elasticity to be true no matter how extreme the
         stretching, but this is unphysical.  The solution to the black
         hole problem may be that the relativity principle, like
         elasticity, is emergent and simply fails at the event horizon.}
\end{figure}

Before explaining how such a thing could be consistent with Einstein
gravity and working out the experimental consequences, let me explain the
basic idea, which is straightforward.  Let us imagine a stretched rubber
sheet with a heavy ball rolling on it, as shown in Fig. 1.  This is the
stripped-down model of gravity familiar from science museum exhibits. The
membrane represents space-time, the ball represents some gravitating
object, the distortion represents the gravitational field, and the motions
of small objects on the membrane represent geodesic trajectories of
satellites.  If the ball is not too heavy then the membrane distorts
elastically to make a slight depression in response to its weight.  Small
objects in the vicinity then fall into this distortion and orbit, membrane
vibrations beamed at the ball scatter, and so forth in analogy with
general relativity. If the ball is too heavy, on the other hand, the
membrane ruptures and the ball falls through.  When this happens the
analogy with Einstein gravity fails completely, since the relativity
principle requires every point in space-time to be locally
indistinguishable from any other, and thus expressly forbids rupture. The
formal statement of this problem is that the black hole event horizon, the
obvious candidate for catastrophic failure in general relativity, exhibits
no singular behavior in any quantity measured in local coordinates.  
However, catastrophic failure is precisely what I suspect is happening at
real black hole surfaces.

What would facilitate this breakdown and at the same time reconcile it
with what we know experimentally about relativity are the limit paradoxes
of continuous quantum phase transitions \cite{stanley, domb}.  Known
physical principles operating at such transitions would enable relativity
to fail quantum mechanically, just as laws of elasticity fail in the
membrane, but so gently that classical Einstein gravity would {\it not} be
violated in any region of space-time experimentally accessible to us
\cite{david}.  This latter point is not obvious, especially if one has not
thought carefully about phases and phase transitions, so one of my main
tasks here today is to explain it convincingly.

The possibility of relativity failure is difficult for most of us to think
about, but should not be. When asked whether relatively applies at the
Planck scale, for example, most physicists will attempt to change the
subject and, in the end, avoid committing one way or the other even
though traditional relativity forbids preferred scales.  Everybody
understands that relativity is believable because it is measured to be
true, not because it ought to be true, and that extrapolating many orders
of magnitude beyond present measurement capability is no less dangerous in
relativity than it is in anything else.

\section{Critical Opalescence}

Let me now begin by reminding you about the simplest known example of a
phase transition, ordinary vaporization.  In 1910 Johannes van der Waals
won the Nobel Prize in physics for his work on non-ideal gasses, and
particularly for his invention of the relation

\begin{equation}
(p + \frac{a}{v^2} )(v - b) = k_B T \; \; \; ,
\label{waals}
\end{equation}

\begin{figure}
\includegraphics{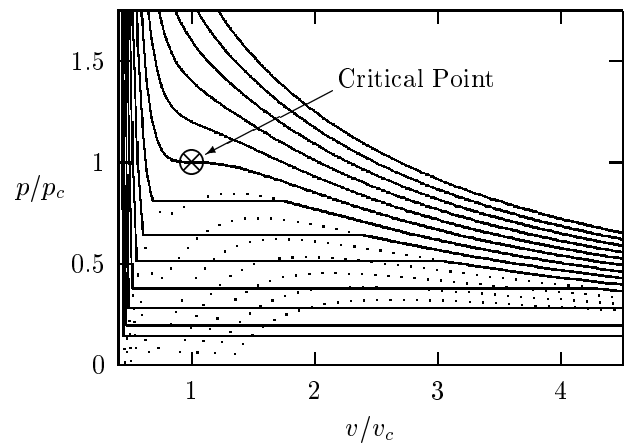}
\caption{Pressure versus volume per atom given by Eq. (\ref{waals})
         for various temperatures.  The critical volume, pressure
         and temperature are $v_c = 3 b$, $p_c = a / 27 b^2$, 
         and $k_B T_c = 8 a/27 b$.  The increment between successive
         isotherms is $\Delta T = 0.05 T_c$. The dotted lines show
         the equation of state before the Maxwell construction.  For
         freon (CCl$_2$F$_2$) the critical temperature is 385 $^\circ$K
         and the critical pressure 4.12 MPa (37 atmospheres).}
\label{dome}
\end{figure}

\noindent
known today as the van der Waals equation of state \cite{maxwell}.  This
approximate description of the non-deal gas captures the essential
features of both departures from ideality in real vapors and the
liquid-vapor transition they anticipate.  $p$ is the pressure, $v$ is the
volume per molecule, T is the temperature, $k_B$ is Boltzmann's constant,
and $a$ and $b$ are parameters characterizing the non-ideality of the
fluid.  $a$ represents the effects of attractive bonding forces between
the molecules, and $b$ represents the volume excluded due to short-range
molecular repulsions. Both parameters are empirically adjusted to fit the
properties of a specific substance.  Setting them to zero produces the
ideal gas law.

The van der Waals equation of state describes the phase transition only
implicitly. It may be seen in Fig. \ref{dome} that its isotherms exhibit
unphysical inflections at low temperatures that result in the bulk modulus

\begin{equation}
k = - v \frac{\partial p}{\partial v}
\end{equation}

\noindent
becoming negative.  This is a symptom of the theory's failure to correctly
describe liquid-vapor coexistence.  From the vast amount of work done on
this problem in the 1970s, culminating in the invention of the Wilson
renormalization group \cite{wilson}, we understand that equations of state
near phase transitions are inherently nonanalytic, and that analytic fits
to them generally produce nonsense when extrapolated across phase
boundaries. These non-analyticities are, however, effects of large size
and disappear when the sample is small.  Taken literally, the Van der
Waals equation of state is a description of a small sample. To apply it to
a large sample we must take into account for the system's tendency to
separate into regions of high and low density.  This is accomplished by
finding a pressure at which the area under a straight line drawn between
the two extremal volumes is the same as that under the inflecting equation
of state.  The end points of this Maxwell construction then define the
liquid and vapor densities \cite{maxwell}.

The critical point - the top of the liquid-vapor dome where the two-phase
region shrinks to zero - is especially important for our discussion of
black holes.  At this point, and this point only, the bulk modulus of the
fluid is identically zero.  For temperatures above the critical
temperature $T_c$, inflection does not occur, the liquid and vapor phases
are physically indistinguishable, and the bulk modulus is positive.  For
temperature less than $T_c$, phase separation occurs, and the bulk modulus
at either end of the Maxwell construction is again positive.  Thus at this
point, and this point only, the speed of sound

\begin{equation}
c = \sqrt{k / \rho} \; \; \; ,
\end{equation}

\noindent
where $M$ and $\rho = M/ v$ are the mass and mass density of the
molecules, vanishes.

In conventional fluids, the vanishing of the sound speed at the critical
point causes critical opalescence, a strange phenomenon in which the fluid
becomes cloudy and opaque to the transmission of light, like an opal.  
This is very dramatic to see.  My colleague Doug Osheroff here at Stanford
has a freshman physics demonstration of critical opalescence that uses
freon as the working fluid.  The critical pressure and temperature of
freon are sufficiently low that one can do this without endangering
students in the front row. Doug sets the temperature to the critical value
and then slowly ramps up the pressure, while Richard Strauss's {\it Also
Sprach Zarathustra} plays in the background.  If he times it right, the
laser shining through the freon winks out at the exact moment the
orchestra plays ``Ta Daa,'' and gets him a standing ovation.

Critical opalescence signals a fundamental failure of hydrodynamics
\cite{domb}.  Equipartition among the degrees of freedom of
compressional sound, which exist by virtue of the principles of
hydrodynamics, requires the departure of the particle density from its
average value to obey

\begin{equation}
< \! \delta \rho ({\bf r}) \delta \rho ({\bf r}') \! >
= \frac{k_B T \; \rho^2}{(2 \pi )^3 k} \int_{|{\bf q} | < 1 / \xi}
e^{i {\bf q} \cdot ( {\bf r} - {\bf r}' )} \; d{\bf q}
\; \; \;  ,
\end{equation}

\noindent
where ${\bf r}$ and ${\bf r}'$ denote different positions in the fluid,
$1/ \xi$ is an ultraviolet cutoff (i.e. a scale at which hydrodynamics
fails), and $< >$ denotes thermal average.  This is the quantity measured
in a light scattering experiment.  It becomes enormous at the critical
point because $k$ goes to zero.  However, since the density correlator
cannot actually become infinite, we know that the key premise of the
calculation, the validity of hydrodynamics, must fail.  This occurs in
practice through the divergence of $\xi$ at the critical point.

The laws of hydrodynamics are emergent.  They are universal, exact
mathematical relationships among measured quantities that develop at long
length scales in liquids and gases.  This development cannot be deduced
from the underlying equations of motion of the atoms. It is a physical
phenomenon - one we know to be exactly true because it is {\it measured}
to be true. Emergent laws are equivalent to, and indistinguishable from,
fundamental laws in all ways but one: they are vulnerable to failure by
simply not emerging.  This is what happens at the critical point.

\begin{figure}
\includegraphics{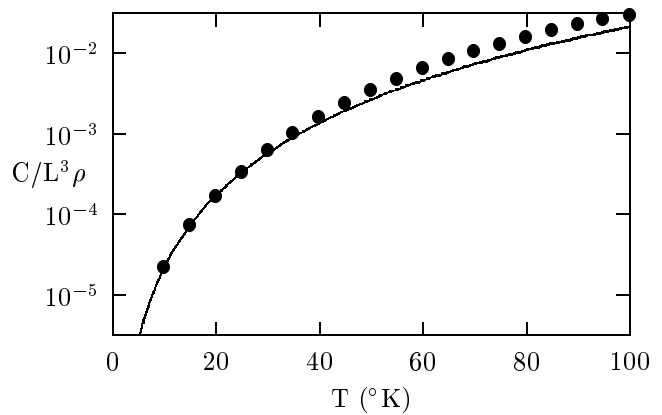}
\caption{Specific heat of Al$_2$O$_3$ in units of calories/gm $^\circ$K
         versus temperature as measured \cite{buyro} (dots)
         and as given by Eq. (\ref{black}) with c = 6142 m/sec.  The
         mass density, bulk modulus and poisson ratio of this material
         are $\rho$ = 3.89 gm/cm$^3$, $k$ = 228 GPa, and $\sigma$ = 0.22,
         respectively. The transverse and longitudinal sound speeds
         computed from these are then $c_t = 6355$ m/sec and $c_\ell =
         10602$ m/sec. Their appropriately weighted average is
         $2c^{-3} = 2c_t^{-3} + c_\ell^{-3}$. }
\label{alumina}
\end{figure}

\section{Quantum Phases}

Phases and phase transitions are not inherently finite-temperature
phenomena.  They occur at zero temperature as well, and are regulated by
principles of emergence in the same way as their finite-temperature
relatives.  The important difference is that zero-temperature phases are
purely quantum-mechanical phenomena.

A quantum phase familiar from everyday experience is the crystalline
solid.  If this is an insulator its low-energy quantum excitations consist
of center-of-mass motion and sound solely. Both are quantum-mechanical.  
The sound waves of the cold solid are quantized ``particles'' with apt
physical similarities with particles of light.  This analogy becomes
increasingly exact as the energy scale is lowered and, in the end, results
in the low-temperature specific heat of all crystalline insulators
becoming the Planck blackbody law

\begin{equation}
\frac{C}{L^3} = \frac{4 \pi^2}{15} (\frac{k_B T}{\hbar c})^3 \; k_B
\label{black}
\end{equation}

\noindent
with the speed of light rescaled down to the speed of sound. This is shown
for the specific case of Al$_2$O$_3$ in Fig. \ref{alumina}.

The liquid and gas phases also exist at zero temperature. The liquid is
realized by either $^4$He nor $^3$He---both of which become superfluids
when cooled to zero temperature (although $^3$He is more complicated)
\cite{glyde}.  The compressional sound waves in these fluids are
quantum-mechanical particles that become more and more accurately defined
and relativistic as the energy scale is lowered.  Their vapor pressures
become unmeasurably small at low temperatures---meaning that both will
puddle at the bottom of a container and will not expand to fill the
available volume.  The vapor is realized by atomic bose-einstein
condensates, the discovery of which by Cornell, Ketterle, and Weiman was
awarded the Nobel Prize in physics in 2001 \cite{cornell,ketterle}.  
Atomic condensates are metastable states of matter and thus not, strictly
speaking, quantum phases.  However, this is unimportant.  They are ground
states of an equivalent fictitious hamiltonian and have all the important
physical properties of phases.  Like $^4$He and $^3$He they exist as
superfluids at ultralow temperatures and have nonzero bulk moduli
\cite{sound}.  Unlike helium, however, they {\it do} expand to fill any
available volume.

The existence of the quantum liquid and gas in nature means that we can
think about the phase transition between them, even though it has never
been observed in the laboratory.  The ideal behavior would be a phase
diagram something like Fig. \ref{dome} except with the ``temperature''
reinterpreted as parameter in the underlying equations of motion.

\begin{figure}
\includegraphics{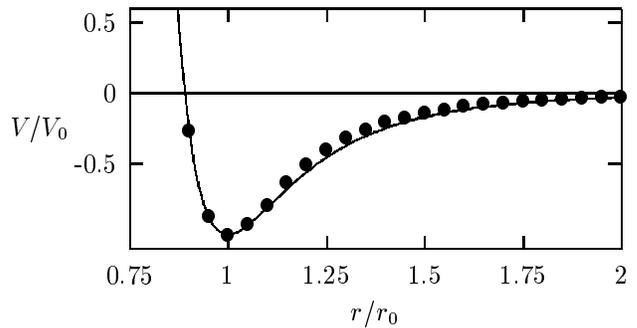}
\caption{Comparison of He-He pair potential measured by atomic-beam 
         scattering \cite{farrar} (dots) with the model of
         Eq. (\ref{vpair}).}
\label{beam}
\end{figure}

Unfortunately, not every parameterization of this transition produces a
phase diagram like that of Fig. \ref{dome}. Real $^4$He is described by
the equations

\begin{equation}
i \hbar \frac{\partial \Psi}{\partial t} = {\cal H} \; \Psi
\label{schroedinger}
\end{equation}

\begin{equation}
{\cal H} = - \sum_j^N \frac{\hbar^2}{2M} \nabla_j^2 + 
V({\bf r}_1 , ... , {\bf r}_N)
\label{hamiltonian}
\end{equation}

\begin{equation}
V({\bf r}_1 , ... , {\bf r}_N ) = \sum_{j < k}
V_{\rm pair} ( {\bf r}_j - {\bf r}_k ) \; \; \; ,
\label{pairsum}
\end{equation}

\noindent
where $V_{\rm pair}$ is a pair potential, given reasonably accurately by

\begin{equation}
V_{\rm pair} ({\bf r}) = V_0 \biggl[ (\frac{r_0}{r})^{12} - 
2 (\frac{r_0}{r})^6 \biggr] \; \; \; ,
\label{vpair}
\end{equation}

\noindent
with $r_0 = 3$ \AA $\;$ and $V_0 = 11 \; ^\circ$K \cite{aziz}.  We know
this to be true because the pair potential has been accurately measured in
atomic beam experiments ({\it cf.} Fig. \ref{beam}).  Also, variational
calculations using this potential predict the correct ground state energy
\cite{mcmillan}, pair correlation function, crystallization
pressure and superfluid transition temperature \cite{ceperley}. Since the
cohesion of the liquid comes from the coefficient of of $1/r^6$ in this
potential, the most obvious way to produce the gas phase is to reduce this
coefficient to zero. However extensive computer modeling in the 1970s
showed that the transition so generated is strongly first-order and cannot
be tuned to criticality with volume \cite{nosanow}.

\section{Superfluidity}

To make a phase diagram like that of Fig. \ref{dome} we must add
multi-body interaction potentials \cite{threeb}.  The simplest way to
accomplish this is to sum short-range components into a density
functional.  Let us consider specifically

\begin{displaymath}
V({\bf r}_1 , ... , {\bf r}_N) = V_1 N + V_2 \sum_{j < k}^N
(\frac{\beta r_0^2}{\pi})^{3/2} e^{- \beta |{\bf r}_j - {\bf r}_k|^2}
+ ...
\end{displaymath}

\begin{equation}
+ V_n \sum_{j_1 < ... < j_n} n^{-3/2} \biggl[
\frac{n \beta r_0^2}{\pi} \biggr]^{(3n-3)/2} \prod_{\mu < \nu}
e^{-\beta |{\bf r}_{j_\mu} - {\bf r}_{j_\nu} |^2} 
\end{equation}

\noindent
where $r_0$ is a characteristic length, in the limit $\beta \rightarrow
\infty$. The multi-center functions in this expression are zero unless $n$
particles are coincident in space, and are normalized to yield 1 when
integrated on all but one of their arguments. With this convention
Eqs. (\ref{schroedinger}) and (\ref{hamiltonian}) may be re-expressed
compactly as the classical lagrangian density \cite{pitaevskii, gross}

\begin{equation}
{\cal L} = \psi^* (i \hbar \frac{\partial}{\partial t} + \mu ) \psi
- \frac{\hbar^2}{2M} | \nabla \psi |^2
- {\cal U} (|\psi({\bf r}|^2) \; \; \; ,
\label{lagrangian}
\end{equation}

\noindent
where $\mu$ is the chemical potential and

\begin{equation}
{\cal U} (|\psi|^2) =
\frac{V_1}{1!} |\psi|^2
+ \frac{V_2}{2!} r_0^3 |\psi|^4
+ \frac{V_3}{3!} r_0^6 |\psi|^6 + ...  \; .
\end{equation}

\noindent
Thus $V_j r_0^{3(j-1)}$ are simply the Taylor expansion coefficients of
${\cal U}$.

The quantum equation of state implicit in models such as Eq.  
(\ref{lagrangian}) are easy to work out when the potentials are weak
\cite{khalatnikov}.  The reason is that the fluid bose condenses, the
operator $\psi$ acquires a vacuum expectation value $\psi \rightarrow <
\!\psi \!>$, and the entire problem becomes classical.  As with any
superfluid order parameter, the square $\psi$ has the physical meaning of
a particle density: $| \psi |^2 = 1/v$.  The usual rules of
canonical quantization then give us for the expected energy

\begin{equation}
< \! {\cal H} \! > = \int \biggl[ \frac{\hbar^2}{2M} | \nabla \psi |^2
+ {\cal U}(|\psi|^2) - \mu | \psi |^2 \biggr] \; d{\bf r}
\; \; \; .
\end{equation}

\noindent
This allows us to identify ${\cal U} ( |\psi|^2)$ as the energy per
unit volume of the fluid as a function of its density, and

\begin{equation}
p = | \psi |^2 {\cal U} ' ( |\psi |^2)
- {\cal U} (| \psi |^2 )
\end{equation}

\noindent
as the pressure. The quantum ground state is implicitly defined by the
energy-minimization condition

\begin{equation}
\mu = {\cal U}' (| \psi |^2 ) \; \; \; .
\end{equation}

It is also easy to work out what ${\cal U}$ must be to give an equation of
state like that of Fig. 2.  We need to modify Eq. (\ref{waals}) somewhat
because the pressure of a quantum fluid is due to interactions and thus
must fall off at least as fast as $1/v^2$ if the potentials are
to be short-ranged. Thus we take

\begin{equation}
(p + \frac{a}{v^4}) (v^2 - b) = d \; \; \; ,
\label{qeos}
\end{equation}

\noindent
where $a$, $b$, and $d$ are hamiltonian parameters.  Since

\begin{equation}
\int p \; dv = \frac{d}{2\sqrt{b}} \ln ( \frac{v - \sqrt{b}}
{v + \sqrt{b}} ) + \frac{a}{3 v^3} \; \; \; ,
\end{equation}

\noindent
we then have

\begin{displaymath}
{\cal U} ( |\psi|^2) = d -  \frac{d}{2 \sqrt{b} | \psi |^2}
\ln ( \frac{ 1 - \sqrt{b} | \psi |^2}{1 + \sqrt{b} | \psi |^2 })
- \frac{a}{3} |\psi |^4
\end{displaymath}

\begin{equation}
= -\frac{a}{3} | \psi |^4 + d \biggl[ \frac{(\sqrt{b} | \psi |^2)^2}{3}
+ \frac{(\sqrt{b} | \psi |^2)^4}{5} + ... \biggr] \; \; \; .
\end{equation}

\noindent
The analysis of this equation of state, including the Maxwell
construction, is exactly the same as with Eq. (\ref{waals}).  It becomes
critical when $d = 8 a /27 b$. 

The dispersion relation of compressional sound implicit in such models has
an important characteristic form. Because it is classical, the motion is
defined by the extremal condition $\delta {\cal L}/\delta \psi^* = 0$,
which is satisfied when

\begin{equation}
i \hbar \frac{\partial \psi}{\partial t} = - \frac{\hbar^2}{2M} \nabla^2
\psi + \biggl[ {\cal U}' (|\psi |^2 ) - \mu \biggr] \psi
\; \; \; .
\end{equation}

\noindent
Assuming that
$\psi = \psi_0 + \delta \psi_R + i \delta \psi_I$, where $\psi_0$ is real
and $\delta \psi_R$ and $\delta \psi_I$ are both small, we have, to
linear order,

\begin{displaymath}
\hbar \frac{\partial (\delta \psi_R)}{\delta t} = - \frac{\hbar^2}{2M}
\nabla^2 (\delta \psi_I )
\end{displaymath}

\begin{equation}
- \hbar \frac{\partial (\delta \psi_I)}{\delta t} = - \frac{\hbar^2}{2M}
\nabla^2 (\delta \psi_R) + 2 {\cal U}'' (\psi_0^2)  \psi_0^2 \;
\delta \psi_R \; \; \; .
\end{equation}

\noindent
Then substituting $\delta \psi_R = \delta \psi_R^{(0)} e^{i ({\bf q} \cdot
{\bf r} - \omega t)}$ and $\delta \psi_I  = \delta \psi_I^{(0)}
e^{i ({\bf q} \cdot {\bf r} - \omega t)}$, and noting that

\begin{equation}
{\cal U}'' (\psi_0^2 ) \; \psi_0^2 = k v \;\; \; ,
\end{equation}

\noindent
we obtain for the dispersion relation

\begin{equation}
\hbar \omega_q = \sqrt{ (\hbar c q)^2 + (\frac{\hbar^2 q^2}{2M})^2}
\; \; \; \; \; \; \; \; \; \; 
( \; c = \sqrt{\frac{k}{\rho}} \; \; )
\; \; \; .
\label{dispersion}
\end{equation}

\begin{figure}
\includegraphics{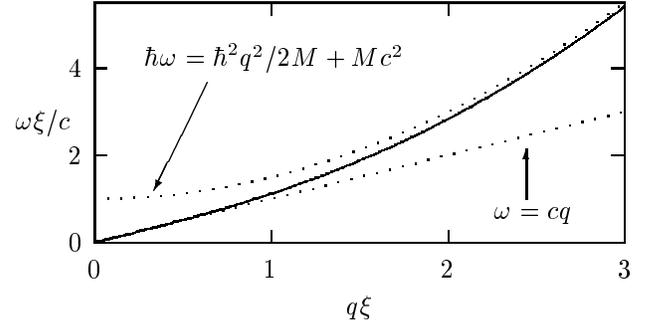}
\caption{Dispersion relation given by Eq. (\ref{dispersion}).
         The dotted lines are asymptotic behaviors at large
         and small $q$. As its momentum increases this excitation evolves
         adiabatically from a relativistic sound quantum into a free
         boson.}
\label{dispersefig}
\end{figure}

\noindent
This is plotted in Fig. \ref{dispersefig}.

Eq. (\ref{dispersion}) contains a length scale

\begin{equation}
\xi = \hbar / Mc \; \; \; .
\end{equation}

\noindent
central to our discussion.  The linear relation $\omega = c q$ expected of
a compressional fluid occurs only when $q \xi << 1$. At scales longer than
$\xi$, the principles of quantum hydrodynamics, the zero-temperature
version of the familiar laws of classical fluids, become exact, and we
obtain a gas of noninteracting relativistic scalar bosons characterized by
velocity $c$. At length scales shorter than $\xi$, on the other hand, the
principles of quantum hydrodynamics fail, and these particles acquire a
decay width and are not guaranteed even to exist.

\section{Quantum Entanglement}

Eq. (\ref{dispersion}) is extremely important for our argument, so let us
derive it by a second method due to Bogoliubov \cite{fetter, bogoliubov}.  
The reasoning in this case is more straightforward but less general, in
that it applies only to the weakly-interacting bose gas.

The starting point of the calculation is again conventional quantum
mechanics as defined by Eqs.  (\ref{schroedinger}) and
(\ref{hamiltonian}), but we use only a pair sum, as in Eq.
(\ref{pairsum}), with the short-range repulsive potential

\begin{equation}
V_{\rm pair} ({\bf r}) = V_2 r_0^3 \; \delta^3 ( {\bf r})
\; \; \; .
\end{equation}

\noindent
With this done Eq. (\ref{hamiltonian}) may be rewritten

\begin{displaymath}
{\cal H} = \sum_{\bf q} \frac{\hbar^2 q^2}{2M} \; a_{\bf q}^\dagger
a_{\bf q}
\end{displaymath}

\begin{equation}
 + \frac{V_2 r_0^3}{2 L^3}
\sum_{\bf q} \sum_{{\bf q}'} \sum_{\Delta {\bf q}}
a_{{\bf q} + \Delta {\bf q}}^\dagger
a_{{\bf q}' - \Delta {\bf q}}^\dagger
a_{{\bf q}'} a_{\bf q} \; \; \; ,
\label{bogham}
\end{equation}

\noindent
where $a_{\bf q}^\dagger$ and $a_{\bf q}$ are bose creation and
annihilation operators satisfying the usual commutation relations

\begin{equation}
[a_{\bf q} , a_{{\bf q}'}] = 0
\; \; \; \; \; \; \; \; \; \;
[ a_{\bf q} , a_{{\bf q}'}^\dagger] = \delta_{{\bf q} {\bf q}'}
\; \; \; .
\end{equation}

\noindent
We assume for simplicity that the bosons live in a box of volume $L^3$
with periodic boundary conditions and that the pair potential is weak.

The calculation is again simplified by the phenomenon of bose
condensation.  This causes the ${\bf q} = 0$ state to acquire macroscopic
occupancy $N$. When this number is thermodynamically large (but not
otherwise) the ${\bf q} = 0$ state becomes a particle reservoir for the
rest of the system. The most significant effect of the potential is to
then to scatter particles out of, and back into, the condensate in pairs.  
The matrix element for this process is always about $N V_2 r_0^3/L^3$,
since $< \! a_0^\dagger a_0 \! > \simeq N$.  Thus if we ignore the
depletion of $N$ due to scattering of bosons out of the condensate then we
may simply replace $a_0$ and $a_0^\dagger$ everywhere they appear in Eq.
(\ref{bogham}) with $\sqrt{N}$. This gives

\begin{displaymath}
{\cal H}_{\rm eff} = \sum_{{\bf q} \neq 0} \biggl\{ ( \frac{\hbar^2
q^2}{2M} + \frac{N V_2 r_0^3}{L^3} ) \; a_{\bf q}^\dagger a_{\bf q}
\end{displaymath}

\begin{equation}
+ \frac{N V_2 r_0^3}{2 L^3} (a_{\bf q}^\dagger a_{-{\bf q}}^\dagger
+ a_{-{\bf q}} a_{\bf q} ) \biggr\} + \frac{N^2 V_2 r_0^3}{2 L^3}
\; \; \; .
\label{boghameff}
\end{equation}

\noindent
${\cal H}_{\rm eff}$ does not conserve particle number, but this is
simply an approximation.  In reality any promotion of particles out of the
condensate will be matched by a corresponding reduction of $N$.

This hamiltonian may be easily diagonialized by canonical transformation.
Let

\begin{equation}
b_{\bf q} = u_q a_{\bf q} + v_q a_{-{\bf q}}^\dagger
\; \; \; .
\end{equation}

\noindent
Then the condition

\begin{equation}
[b_{\bf q} , b_{{\bf q}'}] = 0
\; \; \; \; \; \; \; \; \; \;
[ b_{\bf q} , b_{{\bf q}'}^\dagger] = \delta_{{\bf q} {\bf q}'}
\label{canonical}
\end{equation}

\noindent
requires that $u_q^2 -  v_q^2 = 1$, which is satisfied when

\begin{equation}
u_q = \cosh(\theta_q )
\; \; \; \; \; \; \; \; 
v_q = \sinh(\theta_q )
\; \; \; .
\end{equation}

\noindent
The inverse of Eq. (\ref{canonical}) is $a_{\bf q} = u_q b_{\bf q} -
v_q b_{- {\bf q}}^\dagger$.  Substituting this into Eq. (\ref{boghameff})
we find that the coefficients of $b_{\bf q}^\dagger b_{-{\bf q}}^\dagger$
and $b_{-{\bf q}} b_{\bf q}$ vanish provided that

\begin{equation}
\frac{u_q^2 + v_q^2}{2 v_q u_q}
= \frac{1}{\tanh( 2 \theta_q)}
= 1 + \frac{L^3}{N V_2 r_0^3} \; \frac{\hbar^2 q^2}{2M} 
\; \; \; .
\end{equation}

\noindent
With this choice of $u_q$ and $v_q$ we obtain finally

\begin{displaymath}
{\cal H}_{\rm eff} = \sum_{\bf q} \biggl\{ \hbar \omega_q \;
b_{\bf q}^\dagger b_{\bf q}
+ \frac{\hbar^2 q^2}{2M} v_q^2 - \frac{\hat{N} V_2 r_0^3}{L^3}
u_q v_q \biggr\}
\end{displaymath}

\begin{equation}
+ \frac{\hat{N}V_2 r_0^2}{2 L^3}
\; \; \; \; \; \; \; \; \; 
( \; \hat{N} = N + \sum_{\bf q} v_q^2 \; )
\; ,
\label{bogsepham}
\end{equation}

\noindent
where $\hbar \omega_q$ is given by Eq. (\ref{dispersion}) with

\begin{equation}
M c^2 = \frac{N V_2 r_0^3}{L^3} \; \; \; .
\end{equation}

The ground state implicit in this solution is a highly ``entangled'' state
in which the number of bosons with momentum ${\bf q}$ is correlated with
the number at $- {\bf q}$.  From the condition that every $b_{\bf q}$
annihilate the ground state $| \Psi \! >$ we find that

\begin{equation}
| \Psi \! > = \exp (- \sum_{\bf q} \frac{v_q}{2 u_q} a_{\bf q}^\dagger
a_{- {\bf q}}^\dagger ) (a_0^\dagger)^N | 0 \! > \; \; \; .
\end{equation}

\noindent
The expressions for $u_q$ and $v_q$ may also be obtained by adopting a
ground state of this form as a variational ansatz and minimizing the
expected energy with respect to $v_q/u_q$. They may also be obtained by
minimizing the ground state energy implicit in Eq. (\ref{bogsepham}).

\section{Critical Event Horizon}

Let us now turn to the black hole paradox. Rather than trying to resolve
the problem by promoting a specific theory of gravity, which is probably
not falsifiable anyway, let us use our understanding of quantum fluids
to establish a simple point: Zero-temperature phase transitions generate
the same kinds of apparent illogic one finds at black hole surfaces.  This
result is very general and thus also applies to candidate microscopic
theories of gravity.

\begin{figure}
\includegraphics{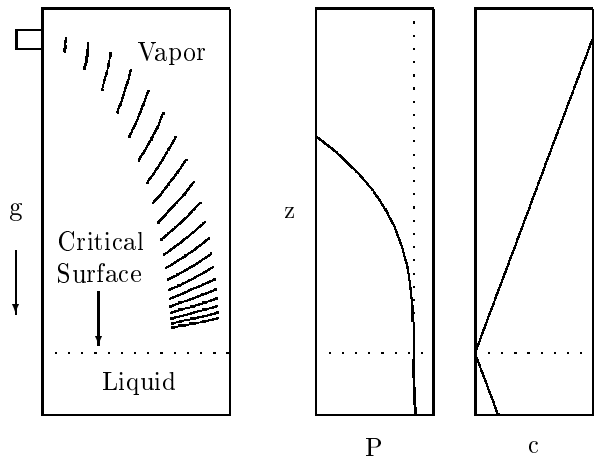}
\caption{Thought experiment in which a tank in earth's gravity is filled
         with cold  superfluid characterized by a critical equation of
        state.  The pressure increases toward the bottom of the tank
        and eventually reaches, and surpasses, the critical value.  Sound
        quanta generated by a transducer are refracted toward this
        ``horizon'' and stall there, just as light does near a classical
        Schwarzschild event horizon.  On the right are the pressure
        profile given by  Eq. (\ref{gravp}) and the sound velocity profile
        given by Eq. (\ref{gravc}).}
\label{gedanken}
\end{figure}

Imagine a thought experiment, illustrated in Fig. \ref{gedanken}, in which
a tall tank on the surface of the earth is filled with a zero-temperature
quantum fluid described by a critical equation of state - for example, Eq.
(\ref{qeos}) with $d = 8a/27b$.  The pressure increases toward the bottom
of the tank due to gravity and, at some critical depth, reaches, and then
surpasses, the critical pressure $p_c = a/27b^2$.  If we now stimulate the
system near the critical surface with a sound transducer, the injected
quanta will be attracted by the surface because the propagation speed is
lower there. This is the exactly same effect as refraction of light toward
the center of a lens or of ocean waves toward a beach---or the
gravitational attraction of light by a black hole.  For both the critical
surface and the black hole horizon this speed actually vanishes, causing
the waves to stall and never reach the surface in finite time. However,
this paradox has a simple resolution in the case of the fluid: The
coherence length $\xi$ diverges to infinity as one approaches the horizon,
and the laws of quantum hydrodynamics fail. The waves cease to have all
meaning as compressional sound and begin doing things disallowed by
hydrodynamics, such as decay and thermalize.

Let us now consider the possibility that this analogy might be {\it apt}
rather than just interesting. Pathologies in the ultraviolet are one of
the central problems of general relativity.  We presently have no way to
resolve them, and the difficulties are so severe that there is serious
talk about a need to change the laws of quantum mechanics \cite{thooft}.  
Moreover the existing experimental record says nothing about this matter
because it does not extend to the Planck scale.  It certainly does not
preclude possibility that relativity might simply fail at the event
horizon through the divergence of a coherence length. Since distinguishing
an emergent phenomenon from a fundamental one at long wavelengths is
impossible, this amounts to a serious logical flaw in the way we normally
think about relativity. It obligates us to take seriously the possibility
that black hole horizons may be phase transitions of the vacuum of
space-time, that they are described inaccurately by the Einstein field
equations in the same way that critical surfaces are inaccurately
described by quantum hydrodynamics, and that elevation of relativity to a
position of transcendence is the source of the entire problem.  Were this
the case, it would instantly resolve all incompatibilities between general
relativity and quantum mechanics, including the unitarity of the
scattering matrix and the loss-of-information paradox.

The analogy between the critical surface and the black hole horizon may be
made quite precise.  In terms of the critical density $\rho_c = M
(3b)^{-1/2}$ we have for the equation of state near criticality

\begin{equation}
\frac{p}{p_c} - 1 = 12 (\frac{\rho}{\rho_c} - 1)^3
\; \; \; .
\end{equation}

\noindent
Since the force of gravity is just a device for achieving a density
increase, we are allowed to weaken gravity at the critical surface 
according to the rule

\begin{equation}
g = g_0 ( 1 - e^{- z^2 / \ell^2}) \; \; \; ,
\end{equation}

\noindent
in order to improve the black hole analogy. We then have

\begin{equation}
\frac{p}{p_c} \simeq 1 - \frac{g_0}{2 \ell^2 c_0^2} z^3
\; \; \; ,
\label{gravp}
\end{equation}

\noindent
where $c_0 = \sqrt{p_c /\rho_c}$ is a velocity scale, and

\begin{equation}
c \simeq  ( \frac{6 c_0 g_0}{\ell^2})^{1/3} |z|
= \frac{|z|}{\tau}
\; \; \; .
\label{gravc}
\end{equation}

\noindent
This is precisely the rule with which the speed of light, measured with
a clock at infinity, vanishes at the event horizon of a Schwarzschild
black hole.  Classical hydrodynamics predicts that small density
fluctuations $\delta \rho$ propagate near the critical surface according
as

\begin{equation}
{\bf \nabla} \cdot [ c^2 {\bf \nabla} (\delta \rho)]
= \frac{\partial^2 (\delta \rho)}{\partial t^2}
\; \; \; .
\end{equation}

\noindent
This is not significantly different from the scalar wave equation

\begin{equation}
{\bf \nabla} \cdot [ c {\bf \nabla} \phi ]
= \frac{1}{c} \frac{\partial^2 \phi }{\partial t^2}
\end{equation}

\noindent
one obtains from

\begin{equation}
\frac{\partial}{\partial x^\mu} ( \sqrt{-g} \; g^{\mu \nu}
\frac{\partial \phi}{\partial x^\nu} ) = 0
\end{equation}

\noindent
using the gravitational metric

\begin{equation}
ds^2 = g_{\mu \nu} dx^\mu dx^\nu = dx^2 + dy^2 + dz^2 - c^ 2 dt^2
\; .
\end{equation}

Failure by means of a diverging coherence length is so subtle that there
is a precise sense in which the failure does not occur at all. The issue
is an order of limits.  Suppose we play a game in which you first declare
how close to the critical surface you wish to go, then I look for a
frequency at which hydrodynamics works all the way down to this height.  
If we do things in this order then I always win, since I can always pick
the frequency sufficiently low that $\omega \xi /c << 1$ on your surface.  
If we do a proper experiment, on the other hand, in which you first fix
the frequency and then I search for the height at which hydrodynamics
fails, then I always lose.  Thus if black hole horizons were like critical
surfaces then there would be a precise sense in which general relativity
was exactly true in all regions of space-time accessible to us.  It would
also be highly misleading.  Our game shows in a physical way that
knowledge of the classical field theory emerging in the $q , \omega
\rightarrow 0$ limit is {\it not} sufficient for predicting all low-energy
things near a phase boundary.  We must also understand the ultraviolet
behavior.  An improper regularization---which is what this improper order
of limits amounts to---can lead to physical nonsense.

\section{Experimental Signatures}

Let us now discuss some experimental properties of the quantum fluid
and its critical surface that might have analogs in quantum gravity
and thus lead to observational tests of these ideas on real black
holes. 

\subsection{Hydrodynamic Failure at High Energies}

The failure of hydrodynamics at the critical surface is a specific case of
a more general effect of ultraviolet breakdown.  Since hydrodynamics is
emergent it must fail in a measurement of the ``vacuum'' far from the
critical surface, either through a deviation from relation $\omega = c q$
or an otherwise disallowed decay, at some characteristic scale $\xi$.  
The properties at this scale must evolve adiabatically to lower energies
as the critical surface is approached, and eventually evolve into the
critical properties.

\subsection{Transparency}

Let us return now to Fig. \ref{gedanken} and consider what happens to the
acoustic energy beamed into the critical region.  Part of the answer is
suggested by Fig. \ref{dispersefig}, which shows that a sound quantum
penetrating into the critical region might morph adiabatically into a
free boson, traverse the region in finite time, and emerge intact from the
other side.  To quantify this effect, however, we must evaluate the rate
at which a boson in the critical region decays. This may be thought
of either as an acoustic nonlinearity or knocking extra bosons out of the
condensate. At the critical point the lagrangian is effectively

\begin{equation}
{\cal L} = \psi^* ( i \hbar \frac{\partial}{\partial t} + \mu ) \psi
- \frac{\hbar^2}{2M} | \nabla \psi |^2 +
3 p_c v_c^2 (|\psi|^2 - \psi_0^2)^4 .
\end{equation}

\noindent
The corresponding quantum hamiltonian is

\begin{displaymath}
{\cal H} = \sum_{\bf q} \frac{\hbar^2 q^2}{2M} a_{\bf q}^\dagger a_{\bf q}
+ \frac{3 p_c v_c^2}{L^3}
\sum_{ {\bf q}_1 {\bf q}_2 {\bf q}_3 {\bf q}_4}
\end{displaymath}

\begin{displaymath}
\times \delta( {\bf q}_1 + {\bf q}_2 + {\bf q}_3 + {\bf q}_4 )
(a_{{\bf q}_1} + a_{-{\bf q}_1}^\dagger)
(a_{{\bf q}_2} + a_{-{\bf q}_2}^\dagger)
\end{displaymath}

\begin{equation}
\times (a_{{\bf q}_3} + a_{-{\bf q}_3}^\dagger)
(a_{{\bf q}_4} + a_{-{\bf q}_4}^\dagger) \; \; \; .
\end{equation}

\noindent
The extremely high order of the nonlinearity means that the fastest
decay process is emission of two extra bosons.  The contribution of this
process to the imaginary part of the boson self-energy is

\begin{displaymath}
{\rm Im} \Sigma_{\bf q}(\omega)
= 12 (\frac{3 p_c v_c^2}{L^3})^2 \sum_{{\bf q}_1
{\bf q}_2} {\rm Im} \biggl[ \hbar \omega
\end{displaymath}

\begin{displaymath}
- \frac{\hbar^2}{2M} (|{\bf q}_1 |^2 + |{\bf q}_2 |^2
+ | {\bf q}_1 + {\bf q}_2 + {\bf q} |^2 ) + i \epsilon \biggr]^{-1}
\end{displaymath}

\begin{equation}
=  - \frac{3\sqrt{3}}{4 \pi^2} (\frac{M}{\hbar^2})^3
(p_c v_c^2)^2 (\hbar \omega -
\frac{\hbar^2 q^2}{6M})^2 \Theta ( \hbar \omega - \frac{\hbar^2 q^2}{6M})
.
\end{equation}

\noindent
Thus the decay rate for a boson of energy $\hbar \omega = \hbar^2 q^2/2M$
is

\begin{equation}
\frac{\hbar}{\tau_{\rm rad}} = \frac{2}{\sqrt{3} \pi^2}
(\frac{M}{\hbar^2})^3 (p_c v_c^2)^2 (\hbar \omega)^2 \; \; \; .
\label{qscatter}
\end{equation}

\noindent
This implies that the free boson becomes more and more sharply defined
as its energy is lowered, so that in the low-energy limit one retrieves
the ideal noninteracting bose gas \cite{sachdev}.

The condition for transparency is $\tau < \tau_{\rm rad}$, where $\tau$ is
defined by Eq. (\ref{gravc}).  The reason is that the time it takes the
boson to traverse the critical region is always $\tau$, regardless of its
energy.  The width of the critical region grows with the boson's momentum
as $\hbar q \tau / M$, but the velocity also grows as $\hbar q/M$, and the
two effects cancel.

\subsection{Critical Opalescence}

If the critical surface is at temperature $T$ then there is a second decay
process in which the boson scatters off of another boson thermally excited
into the vacuum.  This corresponds to conventional critical opalescence.  
It is straightforward to calculate, but we shall just state the result
here.  Up to factors of order 1, the rate as the same as in Eq.
(\ref{qscatter}) except for the substitution $(\hbar \omega)^2 \rightarrow
(k_B T)^{3/2} (\hbar \omega)^{1/2}$.  Thus for frequenceis much larger
than $k_B T/\hbar$ this process is irrelevant.

This has the interesting and important implication that the ratio
$\tau_{\rm rad}/ \tau$ determines whether the critical surface is
optically thin or thick. When this parameter is much less than 1 the
surface is very thin, and therefore not ``black''.

\begin{figure}
\includegraphics{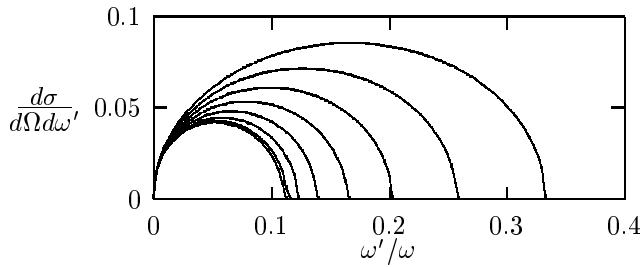}
\caption{Differential cross-section for prompt fluorescence defined by
         Eq. (\ref{crossec}).  The bosons are assumed to be normally
         incident at frequency $\omega$.  The various curves are
         different values of $\theta$ ranging from $0$ to $\pi/2$. 
         The maximum value of $\omega ' / \omega$ for radiation
         returned normally ($\theta = 0$) is 1/9.}
\label{prompt}
\end{figure}

\subsection{Prompt Fluorescence}

There is a net probability of about 0.06 that one of the bosons created in
a decay will escape back out of the surface.  This fluorescence signal is
prompt and has a characteristic spectral shape.  Let us assume for
simplicity that $\tau_{\rm rad} < \tau$, so that all the quanta impinging
on the surface decay. In the opposite case one just reduces the signal by
the decay fraction.  Let us also assume that the bosons are normally
incident at frequency $\omega$. Then the differential cross-section per
unit area $A$ of the surface to scatter back into solid angle $d\Omega$
and between frequency $\omega '$ and $\omega ' + d \omega '$ is

\begin{equation}
\frac{d \sigma}{d\Omega d\omega '} = \frac{27 A}{16 \pi^2 \omega}
\sqrt{3 x[1 - 3 x - 2 \sqrt{x} \cos(\theta)]} \; ,
\label{crossec}
\end{equation}

\noindent
where $x = \omega ' / \omega$ and $\theta$ is the polar angle of exit
just above the critical region, i.e. before the signal is defocused by
outward refraction.  This result is plotted in Fig. \ref{prompt}.

\begin{figure}
\includegraphics{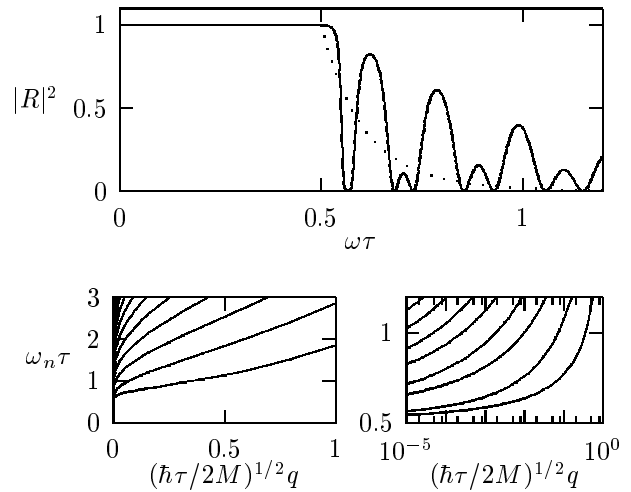}
\caption{Bottom: Resonant frequencies $\omega_n$ obtained by
         solving Eq (\ref{bigwave} ) for various values of
         momentum $q$ in the interfacial plane.  Top:  Reflectivity
         as a function of frequency showing structure at $\omega_n$. 
         The precise shape of this curve depends on how far above the
         surface the law $c = |z|/\tau$ remains valid. The dotted
         curve is a plot of Eq. (\ref{reflect}).}
\label{resonances}
\end{figure}

\subsection{Reflection Resonances}

When $\tau_{\rm rad} >> \tau$ the bosons with parallel momentum $q$ form
bound states in the plane of the critical surface.  This result is
obtained by solving the wave equation

\begin{equation}
\hbar^2 \frac{\partial^2 \psi}{\partial t^2} = (\frac{\hbar}{\tau})^2
{\bf \nabla} \cdot ( z^2 {\bf \nabla} \psi) - (\frac{\hbar^2}
{2M})^2 \nabla^4 \psi
\label{bigwave}
\end{equation}

\noindent
assuming frequency $\omega$ and momentum $\hbar q$ in the plane.
The resulting eigenfrequencies $\omega_n$ are shown in
Fig. \ref{resonances} as a function of $q$. When $q >> (2M/\hbar
\tau)^{1/2}$ the frequencies have the simple harmonic oscillator form

\begin{equation}
\hbar \omega_n \simeq \frac{\hbar^2 q^2}{2M} + ( n + \frac{1}{2} )
\sqrt{2} \; \frac{\hbar}{\tau}
\; \; \; .
\end{equation}

\noindent
This limit is difficult to achieve, for this is the condition implicit in
Eq. (\ref{qscatter}) for $\tau_{\rm rad} < \tau$. When $q$ is very small,
on the other hand, the resonances converge together slowly and, at $q
\rightarrow 0$, collapse to a continuum characterized by the reflectivity

\begin{equation}
| R |^2 = \left[ \begin{array}{lc}
1 & \omega \tau < 1/2 \\
\cosh^{-2}( \pi \sqrt{(\omega \tau)^2 - 1/4} & \omega \tau > 1/2
\end{array} \right]
\label{reflect}
\end{equation} 

\noindent
This is plotted in Fig. \ref{resonances}.  Also plotted is a sketch of
the kind of reflection signal that this effect would tend to
generate - a transmission resonance at every ``bound'' state. More than a
sketch is unfortunately not possible because the details of the
spectrum depend on how far above the surface the relation $c = |z|/\tau$
remains valid.

\subsection{Heat Capacity}

The heat capacity of the critical surface is finite.  Assuming Eq.
(\ref{dispersion}) with $c = |z|/\tau$, we have for the thermal
energy per unit area

\begin{displaymath}
\frac{E}{L^2} = \frac{1}{2 \pi^2} \int_{-\infty}^\infty dz \int_0^\infty
q^2 dq \frac{\hbar \omega_q}{\exp (\beta \hbar \omega_q) - 1}
\end{displaymath}

\begin{equation}
= \frac{\zeta (3)}{\pi} (\frac{k_B T}{\hbar})^3 M \tau
\; \; \; ,
\label{heat}
\end{equation}

\noindent
where $\zeta (3) = 1.202...$.

\section{de Sitter Interior}

In order to talk experimentally about real black holes it is now necessary
for me to speculate about what is inside them.  This is extremely
dangerous since, as we have discussed, it is fundamentally impossible to
infer the nature of one phase from low-energy measurements made on
another.  However, in order to stimulate thinking on what kinds of
measurement one could profitably do on the black hole itself it is
necessary to be concrete, and this requires that we specify what happens
when one crosses the horizon.

\begin{figure}
\includegraphics{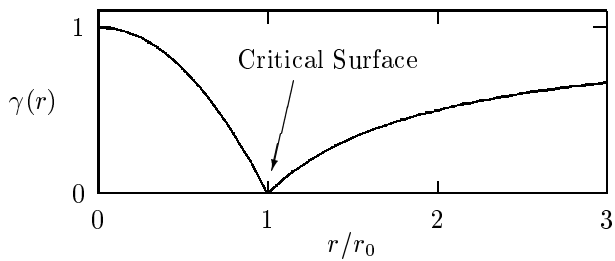}
\caption{Proposed solution for interior of black hole described by 
         Eqs. (\ref{metric}) and (\ref{wild}).  Note that the speed
         of light, measured by clocks at infinity, rises linearly on
         either side of the horizon, just as the sound speed does in
         Fig. \ref{gedanken}. }
\label{desitter}
\end{figure}

I shall presume that that the analogy with Fig. \ref{gedanken} is
literally correct.  Catastrophic jumps of the metric characteristic of a
first-order transition cannot be ruled out on any logical basis, nor can a
non-relativistic equation of state inside the black hole. However, a
continuous transition does the least violence to classical general
relativity and allows the Einstein field equations to be valid, in the
limited sense we have described, everywhere experimentally accessible to
us. If the analogy is apt then the relevant emergent principle - in this
case relativity - must be restored on the other side of the surface in a
mirror-symmetric way. This means that the metric must exist on the other
side, satisfy the Einstein field equations, and be characterized by a
speed of light, measured by clocks at infinity, that {\it increases} with
distance from the horizon, just as the sound speed does on the other side
of the critical surface.  As shown in Fig. \ref{desitter}, this does not
leave one much flexibility.  The boundary conditions at $r = 0$ require
the black hole to contain positive energy density but negative pressure.
This is the characteristic feature of de Sitter space.  Indeed for the the
specific metric

\begin{equation}
ds^2 = \gamma(r) dr^2 + r^2 d\Omega^2 - c^2 \gamma(r) dt^2
\; \; \; ,
\label{metric}
\end{equation}

\noindent
the choice of cosmological-constant matter

\begin{equation}
R_{\mu \nu} - \frac{1}{2} g_{\mu \nu} R = 
\left[ \begin{array}{cc}
3 r_s^{-2} g_{\mu \nu} & r < r_s \\
0 & r > r_s \end{array} \right] \; \; \; ,
\end{equation}

\noindent
where $r_s = 2 G {\cal M}/c^2$, with  ${\cal M}$ the black hole mass,
works nicely, and we obtain

\begin{equation}
\gamma(r) = \left[ \begin{array}{lc}
1 - (r /r_s )^2 & r < r_s \\
1 - r_s / r &  r > r_s
\end{array} \right]
\label{wild}
\end{equation}

\noindent
This leads to the interesting idea that the two phases might be 
distinguished by the values of their cosmological constants.

An important experimental consequence of this solution is that there {\it
would} be a local measurement capable of detecting proximity to a black
hole.  Black hole horizons cannot be phase transitions unless the
relativity principle itself is emergent.  Einstein gravity is based on
very little other than the principles of relativity and equivalence, so if
both of these are true then the predictions of classical general
relativity must also be true---notably that black holes must form in the
conventional way and {\it not} be analogous to the critical surface. Thus
relativity would have to fail at sufficiently high energy scales whether
one were near the black hole or not.  The energy scale of this
failure---the ultraviolet cutoff of the theory---would be both measurable
and position-dependent.  Its lowering to zero would signal proximity to
the event horizon.

A reasonable guess for the scale at which the principles of emergence
should fail is the Planck length $\xi_p = (\hbar G/c^3)^{1/2} = 1.61
\times 10^{-33}$ cm. However, here one must be cautious. The superfluid
analog of Newton's constant is the inverse mass density $\rho^{-1}$.  This
is determined by adding the term

\begin{equation}
\delta {\cal L} = V_2 a^3 \biggl[ | \psi ( {\bf r}  ) |^2
+ | \psi (- {\bf r} ) |^2 \biggr]
\end{equation}

\noindent
to the lagrangian of Eq. (\ref{lagrangian}), and then inducing acoustic
radiation by moving the parameter ${\bf r}(t)$ in a circular orbit of
radius $\ell / 2$ at frequency $\omega$.  After some algebra one finds the
power radiated to be

\begin{equation}
{\cal P} = \frac{1}{5\pi \rho c^5} {\cal M}^2 \ell^4 \omega^8
\; \; \; ,
\end{equation}

\noindent
where ${\cal M}$ is the mass accumulated around ${\bf r}$ by the
potential.  The corresponding quantity in Einstein gravity is

\begin{equation}
{\cal P} = \frac{2}{15} \frac{G}{c^5} {\cal M}^2 \ell^4 \omega^6
\; \; \; .
\end{equation}

\noindent
The disparity in the powers of $\omega$ comes from the fact that
hydrodynamics is a monopolar theory while gravity is quadrupolar.
Combining the mass density and the sound speed dimensionally into a
length, one obtains $(M c /\hbar v)^{1/4} = \eta^{1/4} \xi$, where
$\eta =\xi^3 /v$. Unfortunately the dimensionless constant $\eta$ cannot
be measured in any $q , \omega \rightarrow 0$  experiment.  If $\eta
\simeq 1$, as is the case in superfluid $^4$He, then dimensional analysis
gives a reasonable estimate of $\xi$.  Otherwise it does not.

This model suggests that cosmological black holes may be optically thin,
and thus not ``black'' at all.  For a solar-mass black hole (${\cal M} = 2
\times 10^{33}$ gm) we have [{\it cf.} Eq. (\ref{gravc})] $r_s = 3.0$ km
and $\tau = 2 r_s / c = 2.0 \times 10^{-5}$ sec. To estimate $\tau_{\rm
rad}$ let us assume that the vacuum far from the black hole is analogous
to the quiescent fluid, and that $\eta = 1$ there, so that the coherence
length $\xi$ can be determined by dimensional analysis and thus equals the
Planck length.  Let us further assume that the equivalent fluid density
and pressure are not far from their values at the critical point, so that
$p_c v_c$ is Planck energy $M_p c^2 = \hbar c / \xi$ and $M$ is the Planck
mass $M_p$.  Then we have

\begin{equation}
\tau_{\rm rad} = \frac{\sqrt{3} \pi^2 c}{\xi_p \omega^2}
= \frac{3.2 \times 10^{44} \; {\rm sec}^{-1}}
{\omega^2} \; \; \; ,
\end{equation}

\noindent 
where $\omega$ is the frequency far away from the black hole.  Thus the
horizon would be transparent to gravitons (and presumably any other
particle) of energy less than $\hbar \omega_{\rm max}= 2.6 \times 10^9$
eV.  Were this the case, the black hole would look like a powerful
defocusing lens.

The most important difference between the model of Fig. \ref{desitter}
and a traditional black hole is its finite specific heat.  From Eq.
(\ref{heat}) we find that the total thermal energy contained at the
horizon of a black hole at temperature $T$, measured at infinity, is

\begin{equation}
E = 8 \zeta(3) ( \frac{r_s \; k_B T}
{\hbar})^3 \; \frac{M}{c}\; \; \; .
\label{blacke}
\end{equation}

\noindent 
It is absolutely clear that the cold quantum critical surface does not
radiate, since it is in its ground state, and it is also clear that the
surface may be raised to arbitrary temperatures by adding heat.  Thus this
analogy is fundamentally at odds with Hawking's prediction that a black
hole should emit thermal radiation with a temperature proportional to its
mass\cite{hawking}.  Unruh \cite{unruh} showed a number of years ago that
traditional Hawking radiation is emitted from caustic surfaces of
transsonic superfluid flows, and Jacobson and Volovik \cite{jacobson} have
recently made a good case that this also occurs at ``superluminal''
solitonic domain walls of superfluids.  The long-wavelength description of
these systems is identical to that of the critical surface discussed here,
but the ultraviolet description is different.  Since Hawking's
regularization procedure has no microscopic justification there is reason
for concern that his result may be an artifact of fictitious motion
encoded in the cutoff procedure. The heat capacity implicit in Eq.
(\ref{blacke}) is large. The temperature at which this energy equals
${\cal M}c^2$ for a solar-mass black hole is $k_B T = 141$ eV.

The event horizon in this model contains a large zero-temperature stress,
or negative surface tension, like that in a steel pressure vessel, holding
back the negative pressure of the cosmological-constant matter on the
inside. This, however, is arguably a symptom of the breakdown of
relativity and not physically meaningful stress. It is not detectable in
any measurement performed on the outside other than the the fluorescence
and reflection structure shown in Figs. \ref{prompt} and \ref{resonances},
nor does it exist in any region of space-time where Einstein gravity is
valid (on length scales longer than $\xi$). It resides only on an
infinitely thin surface at which $\xi$ has diverged to infinity and
neither the metric nor the curvature tensor is defined on any scale.  If
one insists on thinking of this stress conventionally then it is large.
For any choice of $\gamma (r)$, conservation of momentum requires the
radial pressure jump across the surface to satisfy

\begin{displaymath}
\Delta T_r^r = \frac{1}{r_0} \int_{r_0 - \epsilon}^{r_0 + \epsilon}
(T_\theta^\theta + T_\phi^\phi - 2 T_r^r) dr
\end{displaymath}

\begin{equation}
= \frac{1}{2 r_0^2} \int_{-\sqrt{r_0 \epsilon}}^{\sqrt{r_0 \epsilon}}
(T_\theta^\theta + T_\phi^\phi - 2 T_z^z) |z| dz
= \frac{3}{4 \pi} \; \frac{{\cal M}}{r_s^3}
\; \; \; .
\end{equation}

\noindent
The thermal contributions to the pressure integral become comparable to
this only when the total thermal energy approaches ${\cal M}
c^2$.  Denoting the these by $\delta T^\mu_\mu$ we have

\begin{equation}
\frac{1}{2 r_0^2} \int
(\delta T_\theta^\theta + \delta T_\phi^\phi ) |z| dz
= \frac{\zeta (3)}{\pi} M ( \frac{ k_B T}{\hbar c})^3
\; \; \; ,
\end{equation}

\noindent
following Eq. (\ref{heat}).

\section{Conclusion}

My lecture today has been intentionally iconoclastic, and I hope you will
all take it in the spirit of fun and as a starting point for reflection
about the gravity problem in new ways. It has been my experience that good
theoretical physics is empowering, in that it enables thinking to take
place that would otherwise not occur, and, in its highest form,
facilitates experiments that would otherwise not be done. This is a
difficult and often dangerous task, as we are paid to be technicians, not
visionaries, and can be just as severely punished for political
incorrectness as a governor or congressman. However, this activity is the
most important thing we do, and perhaps even the {\it only} important
thing we do, for experimentalists are usually smart enough to model for
themselves but cannot take expensive risks without help. For those of you
younger than I am and feeling a bit unsure about how this all works, let
me assure you of one of the great truths of our discipline:  
Experimentalists are amazing and wonderful people.  They have clever
tricks you or I could never guess and are always on the prowl for
something to earn them glory.  Communicating important ideas to them in a
clear and courageous way is the best way I know both to earn one's keep
and to generate science that lasts.

\acknowledgments

The ideas discussed in this manuscript were worked out collaboratively
with my original co-authors George Chapline, Evan Hohlfeld, and David
Santiago \cite{david}, to whom I am most grateful for their friendship,
enthusiasm, and fierce professionalism.  I also gratefully acknowledge
helpful conversations with J. Bjorken, E. Mottola, G. Volovik, and S.-C.
Zhang.

\end{document}